\documentclass[12pt]{article}

\ifx\pdfoutput\undefined
\usepackage[dvips,bookmarks]{hyperref}
\else
\usepackage{hyperref}
\fi
\hypersetup{colorlinks=false,bookmarksopen,bookmarksnumbered,citecolor=blue,
   pdfstartview=FitH}

\usepackage[dvips]{graphicx}
\usepackage{latexsym}

\usepackage{amssymb,amsfonts,amsmath}
\usepackage{graphicx} 
\usepackage{indentfirst}

 \usepackage{bbm}

\parskip=\medskipamount

\def \un{\underline}
\newcommand {\cD}{{\cal D}}

\newcommand {\cN}{{\cal N}}

\def\a{\alpha}
\def \bi{\bibitem}

\def\b{\beta}
\def\c{\chi}
\def\d{\delta}

\def\g{\gamma}
\def\G{\Gamma}

\def\j{\psi}
\def\k{\kappa}
\def\l{\lambda}

\def\q{\theta}
\def\r{\rho}
\def\s{\sigma}

\def\x{\xi}
\def\z{\zeta}

\def\F{\Phi}

\def\P{\Pi}

\def\X{\Xi}

\def\rd{{\rm d}}

\newcommand{\ad}{{\dot{\alpha}}}                           %new
\newcommand{\bd}{{\dot{\beta}}}                            %new
\newcommand{\pa}{\partial}                           %new
\newcommand{\hf}{\frac12}
\newcommand{\vf}{\varphi}
\newcommand{\be}{\begin{equation}}
\newcommand{\ee}{\end{equation}}
\newcommand{\bea}{\begin{eqnarray}}
\newcommand{\eea}{\end{eqnarray}}
\newcommand{\non}{\nonumber}
\def\double #1{#1{\hbox{\kern-2pt $#1$}}}

\renewcommand{\Bar}{\overline}

\begin{document}
%%%%%%%%%%%%%%%%
%%%%%%%%%%%%%%%%
\begin{titlepage}
\begin{flushright}
February, 2010 \\
\end{flushright}
\vspace{5mm}

\begin{center}
{\Large \bf  Variant supercurrent multiplets}
\end{center}

\begin{center}
{\bf
Sergei M. Kuzenko\footnote{kuzenko@cyllene.uwa.edu.au}
} \\
\vspace{5mm}

\footnotesize{
{\it School of Physics M013, The University of Western Australia\\
35 Stirling Highway, Crawley W.A. 6009, Australia}}  
~\\
\vspace{2mm}

\end{center}
\vspace{5mm}

\begin{abstract}
\baselineskip=14pt
In $\cN=1$ rigid supersymmetric theories, there exist three standard realizations
of the supercurrent multiplet corresponding  to the ($i$) old minimal, ($ii$) new minimal and 
($iii$) non-minimal off-shell formulations for $\cN=1$ supergravity.
Recently, Komargodski and Seiberg in  arXiv:1002.2228 put forward a new  supercurrent 
and proved its consistency, although in the past it was believed not to exist.
In this paper, three new variant supercurrent multiplets are proposed.
Implications for supergravity-matter systems are discussed. 
\end{abstract}
\vspace{1cm}

\vfill
\end{titlepage}

\newpage
\renewcommand{\thefootnote}{\arabic{footnote}}
\setcounter{footnote}{0}

\section{Introduction}
\setcounter{equation}{0}
The supercurrent multiplet  \cite{FZ} is a supermultiplet  containing the 
energy-momentum tensor and the supersymmetry current, and therefore 
it is of primary importance in the context of  supersymmetric field theories.
In complete analogy with the energy-momentum tensor, it is fruitful to view the supercurrent 
as the source of supergravity  \cite{OS,FZ2,Siegel}. 
Given a linearized off-shell formulation for $\cN=1$ supergravity, 
the  supercurrent conservation equation can be obtained by coupling  the supergravity 
prepotentials  to external sources and then demanding the resulting action to be invariant under 
the linearized supergravity gauge transformations. One of the prepotentials  is always 
the gravitational superfield $H^{\a \ad}$ \cite{OS} which couples to the supercurrent $J_{\a \ad}$.
The gravitational superfield is accompanied by a superconformal 
compensator. The latter is not universal and  depends on the supergravity formulation chosen.
The source associated with the compensator is sometimes  called a multiplet of anomalies, 
for its components include the trace of the energy momentum tensor and the $\g$-trace 
of the supersymmetry current.

In the literature, there exist three standard supercurrent multiplets 
which correspond to the ($i$) old minimal, ($ii$) new minimal and 
($iii$) non-minimal off-shell formulations for $\cN=1$ supergravity (see, e.g., 
\cite{GGRS} for a  review).
The Ferrara-Zumino supercurrent \cite{FZ} is the most well-known multiplet.
It is characterized by the conservation equation
\bea
{\bar D}^{\ad}J^{(\rm I)}_{\a \ad} = D_\a X~,\qquad {\bar D}_\ad X =0~,
\label{conservation-old}
\eea
and corresponds to the old minimal formulation for $\cN=1$ supergravity \cite{old}
in which the compensator is a chiral scalar $\s$ \cite{S}.
On the other hand, the supercurrent corresponding to the new minimal  supergravity \cite{new}
obeys the conservation law
\bea
{\bar D}^{\ad}{J}^{({\rm II})} _{\a \ad} = { \c}_\a ~,
\qquad {\bar D}_\ad {\c}_\a  =0~, 
\qquad D^\a {\c}_\a = {\bar D}_\ad {\bar {\c}}^\ad~.
\label{conservation-new}
\eea
This equation  reflects, in particular,  the fact that the new minimal compensator,  $G $,
is  real linear  \cite{tensor}. The constraint  on $G$ is solved  \cite{Siegel2}
by introducing a chiral spinor potential $\j_\a$,  by the  rule 
\bea
G= D^\a \j_\a +{\bar D}_\ad {\bar \j}^\ad ~, \qquad {\bar D}_\ad \j_\b =0~.
\label{G1}
\eea
It is defined modulo gauge transformations of the form:
\bea
\d \j_\a = {\rm i}\, {\bar D}^2 D_\a K~, \qquad K = \bar K~.
\label{G2}
\eea
The last equation in (\ref{conservation-new}) is simply the manifestation of this gauge symmetry.
Finally, the supercurrent for the non-minimal supergravity \cite{non-min,SG} is discussed briefly in 
\cite{GGRS}, and it is also reviewed, in a form that differs slightly from that given in \cite{GGRS},  
at the end of section 3. 

Recently Komargodski and Seiberg \cite{KS2}, 
motivated by earlier discussions of the supercurrent multiplets  in theories with Fayet-Iliopoulos terms
\cite{KS,DT,K-FI},   introduced a new supercurrent  with the following conservation law
\bea
{\bar D}^{\ad}{J}^{({\rm IV})} _{\a \ad} = { \c}_\a  +D_\a X~,
\qquad {\bar D}_\ad {\c}_\a  = {\bar D}_\ad X= 0~, 
\qquad D^\a {\c}_\a = {\bar D}_\ad {\bar {\c}}^\ad~.
\label{conservation-IV}
\eea
As pointed out in \cite{KS2}, such a supercurrent had been considered in the past in Ref. \cite{CPS} 
where it had been ruled out as not having a conserved energy-momentum tensor. 
The conclusion of \cite{CPS}  was  shown in \cite{KS2} to be incorrect by explicit component calculations. 
In fact, the consistency of  eq. (\ref{conservation-IV}) follows from the earlier analysis of supercurrents given 
in \cite{K-FI} (see the discussion in section 3 below).

In this note,  three new variant supercurrent multiplets are proposed.
Our consideration is based on the results of \cite{GKP} 
where a classification of off-shell $(3/2,2)$ supermultiplets, or linearized 
supergravity models, is  given. Such models are described in terms of the gravitational superfield
$H_{\un a} := H_{\a \ad}$ and some compensator(s). The latter may occur in one of the following disguises:
({\sl i}) a chiral scalar $\s$,  ${\bar D}_\ad \s=0$; 
({\sl ii}) a  real linear superfield $G$, $\bar G -G ={\bar D}^2 G=0$, which is the gauge-invariant field strength  
of a chiral spinor potential,  eq. (\ref{G1}); 
({\sl iii}) another real linear superfield 
\bea
F= D^\a \r_\a +{\bar D}_\ad {\bar \r}^\ad ~, \qquad {\bar D}_\ad \r_\b =0~,
\label{U1}
\eea
possessing a different supergravity transformation law;
({\sl iv}) a combination of such compensators (say, a complex linear compensator $\G$, 
which emerges in non-minimal supergravity, can be represented as $\G= \s + G +{\rm i} \,F$). 
The linearized supergravity transformations  are:
\vspace{-15pt}
\begin{subequations}
\bea
\label{gauge1}
\d H_{\a \ad} &=& \bar D_{\dot\a}L_{\a}
-D_\a\bar L_{\dot\a}~,  \\
\d\s &=&-\frac 1{12}\bar D^2 D^\a L_\a~, 
\label{gauge2} \\
\d G&=&\frac 14(D^\a\bar D^2 L_\a+\bar D_{\dot\a}D^2 \bar L^{\dot\a})\,\,
\quad \Longrightarrow \quad \d \j_\a = \frac{1}{4} \bar D^2 L_\a ~,
\label{gauge3} \\
\d F&=&\frac{\rm i}{12} (D^\a\bar D^2 L_\a- \bar D_{\dot\a}D^2 \bar L^{\dot\a})
\quad \Longrightarrow \quad \d \r_\a = \frac{\rm i}{12} \bar D^2 L_\a~.
\label{gauge4}
\eea
\end{subequations}
Here the gauge parameter $L_\a$ is an unconstrained spinor superfield.
The analysis carried out in \cite{GKP} results in the following different models 
for linearized supergravity: ({\sl i}) three minimal realizations with $12+12$ off-shell degrees of freedom; 
({\sl ii}) three reducible  realizations with $16+16$ components;
({\sl iii}) one non-minimal formulation with $20+20$ components.
These seven supergravity models lead to different supercurrents.
We  discuss most of these models and associated supercurrents, 
with the exception of  the non-minimal case for which we do not have anything new to say. 
It is useful to formulate the linearized supergravity actions in terms of special
$\cN=1$ superprojectors \cite{Sokatchev,SG2};  
all relevant information about these superprojectors is collected in the Appendix.

\section{Minimal supercurrents}
\setcounter{equation}{0}
It is natural to begin our analysis by considering the supercurrents corresponding 
to the three minimal formulations 
with $12+12$ off-shell degrees of freedom  \cite{GKP}.

The linearized action for old minimal supergravity  is well-known (see, e.g., \cite{GGRS}) and has the form: 
\bea
\label{oldmin}
{\cal S}^{\rm (I)} = \int {\rm d}^8z \, \Big\{ H^{\un a}\Box(-\frac 13\P_{0}^L
+ \frac 12\P_{3/2}^T)H_{\un a} -{\rm i}\,(\s-\bar\s)\pa_{\un a}H^{\un a}-3\s\bar\s \Big\}~.
\eea
We  introduce couplings to external sources, 
\bea
{\cal S}^{\rm (I)} ~\longrightarrow~ 
{\cal S}^{\rm (I)} -\hf \int {\rm d}^8z \,  H^{\a \ad} J_{\a\ad} 
-\frac{3}{2} \Big\{ \int {\rm d}^6z \,  \s X +{\rm c.c.} \Big\} ~,
\eea
and require  invariance under the transformations 
(\ref{gauge1}) and (\ref{gauge2}). Then, it is a two-line calculation to show that $J_{\a \ad} $ and $X$ 
have to obey the equation (\ref{conservation-old}).

Given a chiral scalar $\X$, the supercurrent and the multiplet of anomalies can be transformed as 
\bea
\d J_{\a \ad } =\frac{1}{2} [D_\a , {\bar D}_\ad ] (\X +\bar \X) 
 ={\rm i} \,\pa_{\a \ad} ( \X - \bar \X ) ~, 
\qquad \d X = \frac{1}{4} {\bar D}^2 {\bar \X}~, \qquad {\bar D}_\ad \X =0
\eea
without changing the conservation equation  (\ref{conservation-old}). At the nonlinear supergravity level, 
such an improvement corresponds  (see, e.g., \cite{BK})  to the possibility of adding to the action 
a `non-minimal' term of the form
\bea
\int {\rm d}^8z \, E^{-1} (\X +\bar \X)~,  \qquad {\bar \cD}_\ad \X=0~, 
\eea
which is a generalization of $ R \,\vf^2 $ in  field theory in curved space.

Next, consider the linearized action for new minimal supergravity (see, e.g.,  \cite{BK})
\bea
\label{newmin}
{\cal S}^{\rm (II)}= \int d^8z\,  \Big\{H^{\un a}\Box(-\P_{1/2}^T +\frac 12
\P_{3/2}^T)H_{\un a} + \frac 12 G  [D_\a,\bar D_{\dot\a}]H^{\un a}
+\frac 32 G^2 \Big\}~, 
\eea
where the real linear compensator $G$ should be represented in the form (\ref{G1})   implying 
 gauge invariance (\ref{G2}).
Coupling it to external sources and imposing invariance under the gauge 
transformations (\ref{G2}), (\ref{gauge1}) and (\ref{gauge3}) leads to the supercurrent 
(\ref{conservation-new}).

As is well known (see, e.g., \cite{KS2} for a recent discussion), 
there exists a natural ambiguity  in the definition of ${J}^{({\rm II})} _{\a \ad} $ and $\c_\a$.
Given a U(1) {\it current superfield}, $J$, which is real linear and contains a conserved vector
among its components, the transformation
\bea
\d J^{ ({\rm II}) }_{\a \ad} = \big[ D_\a , {\bar D}_\ad \big] J~, \qquad \d \c_\a =\frac{3}{2} {\bar D}^2 D_\a J~, \quad
\qquad J-\bar J = {\bar D}^2J =0
\eea
preserves the conservation equation  (\ref{conservation-new}). 

The supercurrent (\ref{conservation-new}) can be related to the Ferrara-Zumino one,
eq. (\ref{conservation-old}), 
if the chiral spinor $\c^\a$ can be represented as
\bea
\c_\a = -\frac{1}{4} {\bar D}^2 D_\a V~, \qquad \bar V =V~,
\eea
for some well-defined real scalar $V$. Then we can introduce 
\bea
J^{(\rm I)}_{\a \ad} := J^{(\rm II)}_{\a \ad} +\frac{1}{6} [D_\a , {\bar D}_\ad ] V~, \qquad
X:= -\frac{1}{12} {\bar D}^2V~.
\eea
It is easy to see that  $J^{(\rm I)}_{\a \ad} $ and $X$ obey the conservation equation (\ref{conservation-old}).

There exists one more minimal $12/12$ formulation for linearized supergravity,
which was proposed a few years ago \cite{BGLP}. The corresponding action is 
\bea
\label{2min}
{\cal S}^{\rm (III)} =\int {\rm d}^8z\,  \Big\{H^{\un a}\Box (\frac 13\P_{ 1/2}^L +\frac 12\P_{3/2}^T)
H_{\un a} +F\pa_{\un a}H^{\un a} +\frac 32 F^2 \Big \}~.
\eea
Here $F$ is a real linear superfield that should be treated, similarly to $G$, as the gauge 
invariant field strength of a chiral spinor superfield, eq. (\ref{U1}). 
Coupling this model to external sources and imposing invariance under the gauge 
transformations (\ref{gauge1}) and (\ref{gauge4}), one derives a new supercurrent 
characterized by the conservation equation:
\bea
{\bar D}^{\ad}{J}^{({\rm III})} _{\a \ad} = {\rm i}\, { \eta}_\a ~,
\qquad {\bar D}_\ad {\eta}_\a  =0~, 
\qquad D^\a {\eta}_\a = {\bar D}_\ad {\bar {\eta}}^\ad~.
\label{conservation-novel}
\eea
Here the last equation expresses the fact that the chiral spinor potential associated with  $F$ 
must appear in the action only via the gauge invariant field strength $F$.

In complete analogy with the new minimal supercurrent, 
there is a natural ambiguity  in the definition of ${J}^{({\rm III})} _{\a \ad} $ and $\eta_\a$.
Given a U(1) current superfield $\mathbb J$,  i.e. a real linear superfield,
the transformation
\bea
\d J^{({\rm III})} _{\a \ad} = \pa_{\a \ad} {\mathbb J}, \qquad 
\d \eta_\a =-\frac{1}{4} {\bar D}^2 D_\a {\mathbb J}~, \quad
\qquad {\mathbb J}-\bar {\mathbb J} = {\bar D}^2 {\mathbb J} =0
\eea
preserves the conservation equation  (\ref{conservation-novel}). 

The supercurrent 
(\ref{conservation-novel}) can be related to the Ferrara-Zumino one,
eq. (\ref{conservation-old}), if $\eta_\a$ 
can be  represented in the form:
\bea
\eta_\a = -\frac{1}{4} {\bar D}^2 D_\a {\mathbb V}~, \qquad \bar {\mathbb V} ={\mathbb V}~,
\eea
for some well defined real scalar superfield $\mathbb V$. If we now  define
\bea
J^{(\rm I)}_{\a \ad} := J^{(\rm III)}_{\a \ad} -\pa_{\a\ad} {\mathbb V}~, \qquad
X:= -\frac{\rm i}{4} {\bar D}^2 {\mathbb V}~,
\eea
then $J^{(\rm I)}_{\a \ad} $ and $X$ obey the 
conservation equation (\ref{conservation-old}).

It should be pointed out that the linearized supergravity models 
(\ref{oldmin}), (\ref{newmin}) and (\ref{2min}) are dually equivalent \cite{GKP}.

\section{Reducible supercurrents}
\setcounter{equation}{0}

Let us turn to the derivation of supercurrents corresponding to the three models
with $16+16$ off-shell degrees of freedom \cite{GKP}.
As demonstrated in \cite{GKP}, such theories appear to look like a sum of two of the three
minimal models discussed in the previous section.
Some of these models are linearized versions of  $16/16$ supergravity \cite{GGMW,LLO}
which is known to have no fundamental significance -- it is
just $12/12$ supergravity coupled to matter \cite{Siegel16}.

Consider the type-IV model \cite{GKP}
\bea
\label{hot2}
{\cal S}^{\rm (IV)}&=&\int {\rm d}^8z\, \Big\{H^{\un a}\Box \Big[+8(\a-\frac 1{16})
\P_{0}^L -24(\a-\frac 1{48})\P_{1/2}^T +\frac 12\P_{3/2}^T \Big]H_{\un a}
\nonumber \\
&&-12\Big[(\a-\frac 1{16})(\s+\bar\s) -(\a-\frac 1{48}) G \Big]   [D_\a , \bar D_{\dot\a}]
H^{\un a}
\nonumber \\
&&+72(\a-\frac 1{16})\s\bar\s +36(\a-\frac 1{48})G^2
\Big\}~,~~~~~~~~~~~
\eea
with $\a \neq \frac{1}{16}, \frac{1}{48} $ a real parameter.
This action is invariant under the gauge transformations 
(\ref{gauge1}), (\ref{gauge2}) and (\ref{gauge3}).
If one adds source terms to ${\cal S}^{\rm (IV)}$ 
for all the prepotentials $H^{\a\ad}$, $\s$ and $\j_\a$ 
and demands invariance under the gauge transformations, 
one  immediately  arrives at the  conservation equation (\ref{conservation-IV}).

The operator appearing in the first line of (\ref{hot2}) can be rewritten as 
\bea
 8(\a-\frac 1{16})
\P_{0}^L -24(\a-\frac 1{48})\P_{1/2}^T 
= (\a -\frac{1}{48}) \Big\{ 8\P^L_0 -24 \P^T_{1/2} \Big\} 
-\frac{1}{3} \P^L_0~.
\eea
Using this representation in conjunction with eq. (\ref{A3c}), and also 
setting $\s =0$ in (\ref{hot2}),  one immediately arrives at the linearized action derived in 
subsection 5.2 of \cite{KS2}.

As shown in \cite{KS2}, there is a  freedom in the definition of the triple 
 $(J^{\rm (IV)}_{\a \ad}, \c_\a , X)$ appearing in the conservation equation
 (\ref{conservation-IV}). Given a real scalar $U= \bar U$, 
 the improvement transformation 
 \bea
\d J^{ ({\rm IV}) }_{\a \ad} = \big[ D_\a , {\bar D}_\ad \big] U~, 
\qquad \d \c_\a =\frac{3}{2} {\bar D}^2 D_\a U~, \qquad
\d X = \hf {\bar D}^2U 
\label{improve-IV}
\eea
preserves the defining relations  (\ref{conservation-IV}). 

It may happen that applying a finite transformation (\ref{improve-IV})
results in $\c_\a=0$ or $X=0$, and thus the 
transformed supercurrent is type-I or type-II, respectively. 
This is exactly what happens in the case of the free vector multiplet model 
with a Fayet-Iliopoulos term  studied in \cite{KS,DT,K-FI}.\footnote{In the first version of \cite{DT},
it was claimed that ``no supercurrent supermultiplet exists for globally supersymmetric gauge 
theories with non-zero Fayet-Iliopoulos terms.'' This assertion was shown in \cite{K-FI} 
to be erroneous.  A correct analysis was presented in a revised version of  \cite{DT}.} 
The type-I supercurrent for this model \cite{KS}
\bea
J^{\rm (I)}_{\a \ad} = 2 W_\a {\bar W}_\ad +\frac{2}{3} \x [D_\a, {\bar D}_\ad ] V~, 
\qquad 
X = \frac{1}{3} \x {\bar D}^2 V
\label{JT1} 
\eea 
is not gauge invariant, unlike the type-II supercurrent \cite{K-FI,DT}
\bea
{J}^{\rm (II)}_{\a \ad} = 2 W_\a {\bar W}_\ad ~, 
\qquad 
{\c} _\a = 4 \x W_\a~.
\label{JT2} 
\eea 
Here $W_\a =-\frac{1}{4} {\bar D}^2D_\a V$ is  the chiral field strength
 of a vector multiplet described by the gauge prepotential $V$.
The pairs $(J^{\rm (I)}_{\a \ad} , X)$ and $(J^{\rm (II)}_{\a \ad} , \c_\a)$ given
are related to each other through
a special finite transformation (\ref{improve-IV}) with $U \propto \x V$. 
Applying instead the same finite transformation
but  with a different overall coefficient leads to a type-IV supercurrent. 
Specifically, one can consider the following one-parameter family of supercurrents:
\bea
J^{\rm (IV)}_{\a \ad} = 2 W_\a {\bar W}_\ad +2k  \x [D_\a, {\bar D}_\ad ] V~, \quad 
{\c}^{\rm (IV)} _\a = 4(1-3\k) \x W_\a~,
\quad X^{\rm (IV)} =\k \x  {\bar D}^2 V~, ~~~
\label{JT3}
\eea
with $\k$ a numerical coefficient. The supercurrents  (\ref{JT1}) and (\ref{JT2}) correspond 
to the choices $\k=1/3$ and $\k=0$, respectively. 
Transformation (\ref{JT3}) was essentially behind the analysis in \cite{K-FI}.

Let us turn to the type-V model \cite{GKP}
\bea
\label{hot1}
{\cal S}^{\rm (V)}&=&\int {\rm d}^8z \,\Big\{H^{\un a}\Box \Big[-2(\b-\frac
1{12})\P_{0}^L -2(\b-\frac 14)\P_{1/2}^L+\frac 12\P_{3/2}^T \Big] H_{\un a}
\nonumber \\
&&-6 \Big[ {\rm i}\,(\b-\frac 1{12})(\s-\bar\s)+(\b-\frac 14)F \Big]\pa_{\un a}H^{\un a}
\nonumber \\
&&-18(\b-\frac 1{12})\s\bar\s-9(\b-\frac 14)F^2 \Big\} ~,
\eea
with $\b \neq \frac{1}{4}, \frac{1}{12}$ a real parameter.
This action is invariant under the gauge transformations 
(\ref{gauge1}), (\ref{gauge2}) and (\ref{gauge4}).
It leads to the supercurrent equation
\bea
{\bar D}^{\ad}{J}^{({\rm V})} _{\a \ad} = {\rm i}\,{ \eta}_\a  +D_\a X~,
\qquad {\bar D}_\ad {\eta}_\a  = {\bar D}_\ad X= 0~, 
\qquad D^\a {\eta}_\a = {\bar D}_\ad {\bar {\eta}}^\ad~.
\label{conservation-V}
\eea
Similarly to the situation of the type-IV supercurrent, 
there is a  freedom in the definition of the triple 
 $(J^{\rm (V)}_{\a \ad}, \eta_\a , X)$ appearing in the conservation equation
 (\ref{conservation-IV}). Given a real scalar ${\mathbb U}= \bar {\mathbb U}$, 
the transformation
\bea
\d J^{({\rm V})} _{\a \ad} = \pa_{\a \ad} {\mathbb U}, \qquad 
\d \eta_\a =-\frac{1}{4} {\bar D}^2 D_\a {\mathbb U}~, \quad
\qquad \d X = \frac{\rm i}{4} {\bar D}^2 {\mathbb U} 
\eea
preserves the conservation equation  (\ref{conservation-V}).

It remains to consider the type-VI model \cite{GKP}
\bea
\label{hot3}
{\cal S}^{\rm (VI)}&=&
\int {\rm d}^8z \, \Big\{ H^{\un a}\Box \Big[-2(\g-\frac 14)\P_{1/2}^L
-6(\g-\frac 1{12})\P_{1/2}^T +\frac 12\P_{3/2}^T \Big]H_{\un a}
\nonumber \\
&&+
3(\g-\frac 1{12}) {G} [D_\a,\bar D_{\dot\a}]H^{\un a}
-6(\g-\frac 14) F \pa_{\un a}H^{\un a} \nonumber \\
&&+9(\g-\frac 1{12}) {G}^2 - 9(\g-\frac 14) F^2 \Big\}
~, 
\eea
for a real parameter $\g \neq \frac{1}{4}, \frac{1}{12}$.
This action is invariant under the gauge transformations 
(\ref{gauge1}), (\ref{gauge3}) and (\ref{gauge4}).
It leads to the conservation law
\bea
{\bar D}^{\ad}{J}^{({\rm VI})} _{\a \ad} = { \c}_\a  +{\rm i}\,\eta_\a ~,
\qquad {\bar D}_\ad {\c}_\a  &=& {\bar D}_\ad \eta_\a= 0~, \non \\ 
\qquad D^\a {\c}_\a - {\bar D}_\ad {\bar {\c}}^\ad
&=&D^\a {\eta}_\a - {\bar D}_\ad {\bar {\eta}}^\ad =0
~.
\label{conservation-VI}
\eea
This supercurrent  can be related to the Ferrara-Zumino one,
eq. (\ref{conservation-old}), 
if the chiral spinors $\c_\a$  and $\eta_\a$ are represented as
U(1) field strengths\footnote{Given an unconstrained chiral spinor $\l_\a$, 
${\bar D}_\bd \l_\a= 0$, it can be represented in the form $\l_a =  { \c}_\a  +{\rm i}\,\eta_\a $,
where $\c_\a$ and $\eta_\a$ are given by eqs. (\ref{cs1}) and (\ref{cs2}), respectively.}
\begin{subequations}
\bea
\c_\a &=& -\frac{1}{4} {\bar D}^2 D_\a V~, \qquad \bar V =V~,
\label{cs1} \\
\eta_\a &=& -\frac{1}{4} {\bar D}^2 D_\a {\mathbb V}~, \qquad \bar {\mathbb V} ={\mathbb V}~,
\label{cs2}
\eea
\end{subequations}
for some well-defined real scalars $V$ and $\mathbb V$. Then we can introduce
\bea
J^{(\rm I)}_{\a \ad} := J^{(\rm VI)}_{\a \ad} +\frac{1}{6} [D_\a , {\bar D}_\ad ] V-\pa_{\a\ad} {\mathbb V}
~, \qquad
X:= -\frac{1}{12}{\bar D}^2  (V + 3 {\rm i} \,{\mathbb V})~.
\eea
It is easy to see that  $J^{(\rm I)}_{\a \ad} $ and $X$ obey the 
conservation equation (\ref{conservation-old}).

The conservation law (\ref{conservation-VI}) is,  in fact, related to that corresponding 
to the supercurrent in the non-minimal supergravity (see, e.g., \cite{GGRS}). 
The latter is 
\bea
{\bar D}^{\ad}{J}^{({\rm VII})} _{\a \ad} = -\frac{1}{4} {\bar D}^2 
\z_\a
-\frac{1}{4} \frac{n+1}{3n+1} D_\a 
{\bar D}_\bd {\bar \z}^\bd 
~, \qquad D_{(\a}\z_{\b )} =0~,
\label{conservation-non}
\eea
$n$ is a real parameter, $n\neq-1/3, 0$.\footnote{The constraint on $\z_\a$ in 
(\ref{conservation-non}) is solved by $\z_\a = D_\a W$, for some complex superfield $W$ which is not
always a well-defined local operator.}
Setting here $n=-1$ leads to  (\ref{conservation-VI}).

The models (\ref{hot2}), (\ref{hot1}) and (\ref{hot3}) are equivalent, since they are related to each other
by superfield duality transformations  given in \cite{GKP}.  The real linear superfields $G$ and $F$ 
can be dualized into a chiral scalar and its conjugate. After doing so, one will end up with 
a sum of the old minimal action and that for a free chiral scalar 
(the latter being decoupled from the supergravity prepotentials).
  
\section{Discussion} 
\setcounter{equation}{0}
In this paper we considered six different realizations for the supercurrent multiplet.
All of them are consistent, that is contain  a conserved energy-momentum tensor and 
a supersymmetry current. This follows from the fact that all the multiplets were read off from 
the actions invariant under linearized supergravity transformations, eqs. (\ref{gauge1})--(\ref{gauge4}),
 generated by an unconstrained parameter $L_\a (x,\q, \q)$; the linearized general coordinate 
 and local supersymmetry transformations are part of the gauge freedom.
In other words, there  is no need to carry out a component analysis of 
the supercurrent in order to check that the energy-momentum tensor and 
the supersymmetry current are conserved. 

The type-III supergravity formulation, eq. (\ref{2min}), possesses quite interesting properties 
\cite{GKP}. However, its extension beyond  the linearized approximation is not known.
This means that the supercurrent multiplets $(J^{\rm (III)}_{\a \ad}, \eta_\a )$, 
$(J^{\rm (V)}_{\a \ad}, \eta_\a , X)$ and $(J^{\rm (VI)}_{\a \ad}, \eta_\a , \c_\a)$
are of purely academic interest, at least at present. 

 Komargodski and Seiberg \cite{KS2} demonstrated that there exist interesting 
 supersymmetric theories\footnote{Such theories include ({\sl i})
 $\cN=1$ nonlinear sigma-models with a non-exact K\"ahler form; ({\sl ii}) models with
  Fayet-Iliopoulos terms.} 
 for which the Ferrara-Zumino supercurrent (\ref{conservation-old}) is not well defined.
 They also showed that the type-IV supercurrent, eq. (\ref{conservation-IV}), always exists. 
 Does that mean that it is necessary to develop an off-shell supergravity formulation 
 that automatically leads to the type-IV supercurrent? In our opinion, the answer is {\it no}. 
 It is well known that  any $\cN=1$ supergravity-matter system (including the new minimal 
 and non-minimal supergravity theories) can be realized as a coupling of 
 the old minimal supergravity to matter \cite{FGKV,BK}. 
 Keeping in mind this general result, and the fact that the supercurrent multiplet is the source 
 of supergravity, it is more appropriate 
 to re-formulate the conclusion of \cite{KS2} in a more positive form: the Ferrara-Zumino supercurrent 
 (\ref{conservation-old}) always exists, modulo an improvement transformation of the form:
 \bea
 J^{ ({\rm I}) }_{\a \ad} ~\to ~ J^{ ({\rm I}) }_{\a \ad}+
\big[ D_\a , {\bar D}_\ad \big] U~, 
\qquad 
X~\to~ X + \hf {\bar D}^2U ~, \qquad \bar U =U~.
\eea
Such an improvement results in the conservation law  (\ref{conservation-IV}),
in which $\c_\a =\frac{3}{2} {\bar D}^2 D_\a U$.
\\

\noindent
{\bf Acknowledgements:}\\
The author is grateful to Joseph Novak for comments on the manuscript.
This work  is supported in part by the Australian Research Council and the Australian Academy of Science.
\\

\noindent
{\bf Comments added:} \\
There still remains an additional possibility to generate a reducible supercurrent multiplet 
that has not been considered above.\footnote{This option was  used  in \cite{GKP}
as a means to introduce the linearized non-minimal supergravity. The authors of \cite{GKP}
considered a gauge-invariant action of the form 
$\a \,{\cal S}^{\rm (I)} +\b\,{\cal S}^{\rm (II)} +\g \,{\cal S}^{\rm (III)}$, with $\a +\b +\g=1$,
and showed that its dependence on the compensators $\s$, $G$ and $F$ 
occurs only via a complex linear superfield $\G =a \s +b G + {\rm i} \,cF$ and its conjugate $\bar \G$, 
for some real coefficients $a,b,c$. As a functional of $H_{\a \ad}$, $\G$ and $\bar \G$, 
the action describes linearized non-minimal supergravity parametrized by a complex parameter
$n$.}
Specifically, one can start from a (two-parameter) sum 
of the three minimal models ${\cal S}^{\rm (I)}$, ${\cal S}^{\rm (II)}$
and ${\cal S}^{\rm (III)}$, and use it to read off the corresponding supercurrent. 
One then ends up with the conservation law
\bea
{\bar D}^{\ad}{J}^{({\rm VIII})} _{\a \ad} = { \c}_\a  +{\rm i}\,\eta_\a +D_\a {\mathbb X}~,
\qquad {\bar D}_\ad {\c}_\a  &=& {\bar D}_\ad \eta_\a= {\bar D}_\ad {\mathbb X}=0~, \non \\ 
\qquad D^\a {\c}_\a - {\bar D}_\ad {\bar {\c}}^\ad
&=&D^\a {\eta}_\a - {\bar D}_\ad {\bar {\eta}}^\ad = 0~.
\label{conservation-VIII}
\eea
This supercurrent embraces the previously considered six multiplets as special cases.

The supercurrent   $(J^{\rm (VIII)}_{\a \ad}, \c_\a , \eta_\a,X)$ proves to be equivalent to that derived 
eight years ago\footnote{The author is grateful to Ivo Sachs for reminding him of  \cite{MSW}.}
by Magro, Sachs and Wolf  \cite{MSW}, with the aid of their superfield Noether procedure (see also \cite{Osborn}), 
{\it provided} the chiral spinors $\c_\a$  and $\eta_\a$ can be
  represented as U(1) field strengths, eqs. (\ref{cs1}) and (\ref{cs2}),  
associated with  globally well-defined
scalar prepotentials $V$ and $\mathbb V$. 
However, the resulting supercurrent
\bea
{\bar D}^{\ad}{J}^{({\rm VIII})} _{\a \ad} =
 -\frac{1}{4} {\bar D}^2 D_\a (V +{\rm i}{\mathbb V}) +D_\a {\mathbb X}~, \qquad
{\bar V} - V= {\bar {\mathbb V}} -{\mathbb V} = {\bar D}_\ad {\mathbb X}=0~,
 \eea
 which is the most general supercurrent given in  \cite{MSW},
 is obviously equivalent to the Ferrara-Zumino one, eq. (\ref{conservation-old}).
Indeed, our earlier consideration shows  that we can introduce
\bea
J^{(\rm I)}_{\a \ad} := J^{(\rm VIII)}_{\a \ad} +\frac{1}{6} [D_\a , {\bar D}_\ad ] V-\pa_{\a\ad} {\mathbb V}
~, \qquad
X:= {\mathbb X} -\frac{1}{12}{\bar D}^2  (V + 3 {\rm i} \,{\mathbb V})~,
\eea
where $J^{(\rm I)}_{\a \ad} $ and $X$ obey the 
conservation equation (\ref{conservation-old}).
On the other hand, from  the work of \cite{KS2}  we know of the existence of nontrivial supersymmetric theories
for which the Ferrara-Zumino supercurrent (\ref{conservation-old}) is not well defined.
In particular, this  takes place in the case of  supersymmetric 
nonlinear sigma-models
\bea
S[\F, \bar \F] =  \int 
\rd^8 z\,
 K(\Phi^{I},
 {\bar \Phi}{}^{\bar{J}})  +  \left\{ \int {\rm d}^6z \,  W(\F^I) +{\rm c.c.} \right\} 
\label{nact4}
\eea
for which the K\"aher two-form of the target space is not exact. 
Then, the type-IV triplet $(J^{\rm (IV)}_{\a \ad}, \c_\a , X)$
involves the well-defiined local operators \cite{KS2}
\bea 
J^{\rm (IV)}_{\a \ad} = ({\bar D}_\ad {\bar \F}^{\bar J}) (D_\a \F^I) K_{I\bar J}~, \qquad 
\c_\a =-\hf {\bar D}^2 D_\a K~, \qquad X =-2 W~
\eea
which are invariant under K\"ahler transformations. 
Unlike $\c_\a$, however, its prepotential $V= 2K (\F , \bar \F)$ is not globally well-defined.
It would be important to understand whether the superfield Noether procedure of \cite{MSW,Osborn} 
is flexible enough to account for such exotic models.

\appendix

\section{Superprojectors} 
\setcounter{equation}{0}

The gravitational superfield can be represented as a superposition of SUSY irreducible 
components,
\bea
 H_{\un a} =  \Big(\P^L_{0}+\P^L_{1/2} +\P^T_{1}+\P^T_{1/2}+\P^T_{3/2} 
\Big)H_{\un a}~,
\eea
by making use of the relevant superprojectors
\cite{SG2,GKP}
\begin{subequations}
\bea
\P^L_{0}H_{\un a}&=&-{\frac 1{32}}\Box^{{}_{-2}}\pa_{\un a}\{ D^2,
\bar D^2 \}\pa_{\un c}H^{\un c} ~,
\\
\P^L_{1/2}H_{\un a}&=&
{\frac 1{16}}\Box^{{}_{-2}}\pa_{\un a}D^\d \bar D^2D_\d \pa_{\un
c}H^{\un c}~, \\
\P^T_{1/2}H_{\un a}&=&{\frac 1{3!8}}\Box^{{}_{-2}}\pa_{\dot\a}^{~\b}
[D_\b \bar D^2D^\d \pa_{(\a}^{~~\dot\b}H_{\d)\dot\b}+D_\a\bar D^2D^\d
\pa_{(\b}^{~~\dot\b}H_{\d)\dot\b}]~, \\
\P^T_{1}H_{\un a}&=&{\frac 1{32}}\Box^{{}_{-2}}\pa_{\dot\a}^{~\b}\{
D^2, \bar D^2\} \pa_{(\a}^{~~\dot\b}H_{\b)\dot\b} ~,\\
\P^T_{3/2}H_{\un a}&=&
-{\frac 1{3!8}}\Box^{{}_{-2}}\pa_{\dot\a}^{~\b} D^\g\bar D^2
D_{(\g}\pa_\a^{~\dot\b}H_{\b)\dot\b}~.
\eea
\end{subequations}
Here the superscripts $L$ and $T$ denote longitudinal and transverse
projectors, while the subscripts $0, 1/2, 1, 3/2$ stand for  superspin.
One can readily express the action in terms of the superprojectors.
It is a  D-algebra exercise to show
\begin{subequations}
\bea
D^\g\bar D^2D_\g H_{\un a} &=&
-8 \Box (\P^L_{1/2}+\P^T_{1/2}
+\P^T_{3/2})H_{\un a}~,\\
\pa_{\un a}\pa^{\un b}H_{\un b}&=&
-2\Box ( \P^L_{0} +\P^L_{1/2})H_{\un a}~, \\
\label{doubletrouble}
[D_\a ,\bar D_{\dot\a}][D_\b ,\Bar D_{\dot\b}]H^{\un  b}&=&
+ \Box (8\P_{0}^{L} -24\,\P^T_{1/2}) H_{\un a}~.
\label{A3c}
\eea
\end{subequations}

\small{

}


\begin{thebibliography}{66}

%\cite{Ferrara:1974pz}
\bibitem{FZ}
  S.~Ferrara and B.~Zumino,
``Transformation properties of the supercurrent,''
  Nucl.\ Phys.\  B {\bf 87}, 207 (1975).
  %%CITATION = NUPHA,B87,207;%%
  

%\cite{Ogievetsky:1976qc}
\bibitem{OS}
V.~Ogievetsky and E.~Sokatchev,
``On vector superfield generated by supercurrent,''
Nucl.\ Phys.\  B {\bf 124}, 309 (1977).
  %%CITATION = NUPHA,B124,309;%%

%\cite{Ferrara:1977mv}
\bibitem{FZ2}
  S.~Ferrara and B.~Zumino,
  ``Structure of conformal supergravity,''
  Nucl.\ Phys.\  B {\bf 134}, 301 (1978).
  %%CITATION = NUPHA,B134,301;%%


%\cite{Siegel:1977km}
\bibitem{Siegel}
  W.~Siegel,
  ``A derivation of the supercurrent superfield,''
Harvard  preprint  HUTP-77/A089 (December, 1977). 
  %%CITATION = HUTP-77/A089;%%
  
\bibitem{GGRS}
S.~J.~Gates Jr., M.~T.~Grisaru, M.~Ro\v{c}ek and W.~Siegel,
{\it Superspace, or One Thousand 
and One Lessons in Supersymmetry},
Benjamin/Cummings (Reading, MA),  1983, hep-th/0108200.

\bi{old}
J.~Wess and B.~Zumino,
``Superfield Lagrangian for supergravity,''
Phys.\ Lett.\  B {\bf 74}, 51 (1978);
%%CITATION = PHLTA,B74,51;%%
K.~S.~Stelle and P.~C.~West,
``Minimal auxiliary fields for supergravity,''
Phys.\ Lett.\  B {\bf 74},  330 (1978);
S.~Ferrara and P.~van Nieuwenhuizen,
``The auxiliary fields of supergravity,''
Phys.\ Lett.\  B {\bf 74}, 333 (1978).
%%CITATION = PHLTA,B74,333;%%

\bibitem{S}
W.~Siegel,
``Solution to constraints in Wess-Zumino supergravity formalism,''
Nucl.\ Phys.\  B {\bf 142}, 301 (1978). 

%\bibitem{SG}
%W.~Siegel and S.~J.~Gates Jr.
 %``Superfield supergravity,''  Nucl.\ Phys.\  B {\bf 147}, 77 (1979).
  
\bibitem{new}
V.~P.~Akulov, D.~V.~Volkov and V.~A.~Soroka,
``Generally covariant theories of gauge fields on superspace,'' 
Theor.\ Math.\ Phys.\  {\bf 31}, 285 (1977);
M.~F.~Sohnius and P.~C.~West,
``An alternative minimal off-shell version of N=1 supergravity,''
Phys.\ Lett.\  B {\bf 105}, 353 (1981).

%\cite{Bedding:1981na}
\bibitem{tensor}
  S.~P.~Bedding and W.~Lang,
  ``Linearized superfield formulation of the new minimal N=1 supergravity,''
  Nucl.\ Phys.\  B {\bf 196}, 532 (1982);
  %%CITATION = NUPHA,B196,532;%%
  P.~S.~Howe, K.~S.~Stelle and P.~K.~Townsend,
  ``The vanishing volume of N=1 superspace,''
  Phys.\ Lett.\  B {\bf 107}, 420 (1981);
  S.~J.~Gates Jr., M.~Ro\v{c}ek and W.~Siegel,
  ``Solution to constraints for n=0 supergravity,''
  Nucl.\ Phys.\  B {\bf 198}, 113 (1982).
  %%CITATION = NUPHA,B198,113;%%


%\cite{Siegel:1979ai}
\bibitem{Siegel2}
W.~Siegel,
``Gauge spinor superfield as a scalar multiplet,''
Phys.\ Lett.\  B {\bf 85}, 333 (1979).
  %%CITATION = PHLTA,B85,333;%%

\bi{non-min}
P.~Breitenlohner,
``Some invariant Lagrangians for local supersymmetry,''
Nucl.\ Phys.\ {\bf B124}, 500 (1977).


\bibitem{SG}
W.~Siegel and S.~J.~Gates Jr.
 ``Superfield supergravity,''  Nucl.\ Phys.\  B {\bf 147}, 77 (1979).
  

%\cite{Komargodski:2010rb}
\bibitem{KS2}
Z.~Komargodski and N.~Seiberg,
  ``Comments on supercurrent multiplets, supersymmetric field theories and
  supergravity,''
  arXiv:1002.2228 [hep-th].
  %%CITATION = ARXIV:1002.2228;%%

%\cite{Komargodski:2009pc}
\bibitem{KS}
  Z.~Komargodski and N.~Seiberg,
``Comments on the Fayet-Iliopoulos term in field theory and supergravity,''
 JHEP {\bf 0906}, 007 (2009)  [arXiv:0904.1159 [hep-th]].
  %%CITATION = JHEPA,0906,007;%%

%\cite{Dienes:2009td}
\bibitem{DT}
  K.~R.~Dienes and B.~Thomas,
  ``On the inconsistency of Fayet-Iliopoulos terms in supergravity theories,''
  arXiv:0911.0677v2 [hep-th].
  %%CITATION = ARXIV:0911.0677;%%


%\cite{Kuzenko:2009ym}
\bibitem{K-FI}
  S.~M.~Kuzenko,
  ``The Fayet-Iliopoulos term and nonlinear self-duality,''
  arXiv:0911.5190 [hep-th].
  %%CITATION = ARXIV:0911.5190;%%

%\cite{Clark:1978jx}
\bibitem{CPS}
  T.~E.~Clark, O.~Piguet and K.~Sibold,
  ``Supercurrents, renormalization and anomalies,''
  Nucl.\ Phys.\  B {\bf 143}, 445 (1978).
  %%CITATION = NUPHA,B143,445;%%


%\cite{Gates:2003cz}
\bibitem{GKP}
  S.~J.~Gates Jr., S.~M.~Kuzenko and J.~Phillips,
  ``The off-shell (3/2,2) supermultiplets revisited,''
  Phys.\ Lett.\  B {\bf 576}, 97 (2003)
  [arXiv:hep-th/0306288].
  %%CITATION = PHLTA,B576,97;%%

\bibitem{Sokatchev}
E.~Sokatchev,  ``Projection operators  and supplementary conditions for 
superfields with arbitrary spin,''
Nucl.\ Phys.\ B {\bf 99}, 96 (1975).  

%\cite{Siegel:1981ec}
\bibitem{SG2}
W.~Siegel and S.~J.~Gates  Jr.,  ``Superprojectors,''
Nucl.\ Phys.\  B {\bf 189}, 295 (1981).
  %%CITATION = NUPHA,B189,295;%%
 
\bibitem{BK} I.~L.~Buchbinder and S.~M.~Kuzenko,
{\it Ideas and Methods of Supersymmetry and
Supergravity or a Walk Through Superspace},
IOP, Bristol, 1998. 
 
%\cite{Buchbinder:2002gh}
\bibitem{BGLP}
  I.~L.~Buchbinder, S.~J.~Gates Jr., W.~D.~Linch and J.~Phillips,
  ``New 4D, N = 1 superfield theory: Model of free massive superspin-3/2
  multiplet,''
  Phys.\ Lett.\  B {\bf 535}, 280 (2002)
  [arXiv:hep-th/0201096].
  %%CITATION = PHLTA,B535,280;%%


%\cite{Girardi:1984vq}
\bibitem{GGMW}
  G.~Girardi, R.~Grimm, M.~M\"uller and J.~Wess,
  ``Antisymmetric tensor gauge potential in curved superspace and a (16+16)
  supergravity multiplet,''
  Phys.\ Lett.\  B {\bf 147}, 81 (1984).
  %%CITATION = PHLTA,B147,81;%%

%\cite{Lang:1985xk}
\bibitem{LLO}
  W.~Lang, J.~Louis and B.~A.~Ovrut,
  ``(16+16) supergravity coupled to matter: The low-energy limit of the
  superstring,''
  Phys.\ Lett.\  B {\bf 158}, 40 (1985).
  %%CITATION = PHLTA,B158,40;%%

%\cite{Siegel:1986sv}
\bibitem{Siegel16}
  W.~Siegel,
  ``16/16 supergravity,''
  Class.\ Quant.\ Grav.\  {\bf 3} (1986) L47.
  %%CITATION = CQGRD,3,L47;%%


%\cite{Ferrara:1983dh}
\bibitem{FGKV}
  S.~Ferrara, L.~Girardello, T.~Kugo and A.~Van Proeyen,
  ``Relation between different auxiliary field formulations of N=1 supergravity
  coupled to matter,''
  Nucl.\ Phys.\  B {\bf 223}, 191 (1983).
  %%CITATION = NUPHA,B223,191;%%


%\cite{Magro:2001aj}
\bibitem{MSW}
  M.~Magro, I.~Sachs and S.~Wolf,
  ``Superfield Noether procedure,''
  Annals Phys.\  {\bf 298}, 123 (2002)
  [arXiv:hep-th/0110131].
  %%CITATION = APNYA,298,123;%%

%\cite{Osborn:1998qu}
\bibitem{Osborn}
  H.~Osborn,
  ``N = 1 superconformal symmetry in four-dimensional quantum field theory,''
  Annals Phys.\  {\bf 272}, 243 (1999)
  [arXiv:hep-th/9808041].
  %%CITATION = APNYA,272,243;%%




\end{thebibliography}
\end{document}